\documentclass[twocolumn,showpacs,aps,epsfig]{revtex4}

\usepackage{graphicx}
\usepackage{epstopdf}
\usepackage[center]{subfigure}

\begin{document}

%%%%%%%%%%%%%%%%%%%%%%%%%%%%%%%%%%%%%%%%%%%%%%%%%%%%%%%%%%%%%%%
 \newcommand{\bq}{\begin{equation}}
 \newcommand{\eq}{\end{equation}}
 \newcommand{\bqn}{\begin{eqnarray}}
 \newcommand{\eqn}{\end{eqnarray}}
 \newcommand{\nb}{\nonumber}
 \newcommand{\lb}{\label}
 %%%%%%%%%%%%%%%%%%%%%%%%%%%%%%%%%%%%%%%%%%%%%%%%%%%%%%%%%%%%%%%

\title{Existence of black holes in Friedmann-Robertson-Walker universe dominated
by dark energy}
\author{Zhong-Heng Li ${ }^{1, 2, 3}$ \thanks {Email: Zhong-Heng$\_$@baylor.edu}
and Anzhong Wang ${ }^{2, 3}$ \thanks {Email: Anzhong$\_$Wang@baylor.edu}
\thanks{Electronic address:  Anzhong$\_$Wang@baylor.edu}  }
\affiliation{ ${ }^{1}$ Department of Physics, Zhanjiang Normal University,
Zhanjiang 524048, China\\
${ }^{2}$ Department of Physics, Zhejiang University
of Technology, Hangzhou 310032, China\\
${ }^{3}$ CASPER, Department of Physics, Baylor University, Waco,
Texas 76798, USA}
\date{\today}

\begin{abstract}

We study the existence of black holes in a homogeneous and isotropic
expanding Friedmann-Robertson-Walker (FRW) universe dominated by dark 
energy. We show that black holes can exist  in such a universe by 
considering some specific McVittie solutions. Although  these solutions
violate all three energy conditions, the FRW background does  satisfy
the weak energy condition.

\end{abstract}

\vspace{.7cm} 

\maketitle

%\vspace{1.cm}

\section{Introduction}

\renewcommand{\theequation}{1.\arabic{equation}}
\setcounter{equation}{0}

Astronomical observations indicate that about $2/3$ of total energy
of the universe is to be attributed to a fluid with equation of
state $w<-1/3$, which drives the acceleration of universe, usually
called dark energy \cite{1}. Up to now, numerous cosmological models with
dark energy have been proposed. Another fundamental physical issue
is how dark energy affects the formation and evolution of black
holes. Babichev {\em et al} \cite{2} considered the 
 accretion of  a relativistic perfect fluid
onto  black holes and showed that, if the expansion of the
universe is dominated by phantom energy, black holes will decrease
their mass due to phantom energy accretion, and tend to zero at the
time of the big rip. Harada {\em et al} \cite{3} concluded that there is
no self-similar black hole solution in a universe with a stiff fluid
or scalar field or quintessence. Even some asserted that black
holes might not exist in our real world, because the large negative pressure
might prevent black holes from forming \cite{4}.

On the other hand, Cai and Wang \cite{5}  investigated the black hole
formation from collapsing dust in the background of dark energy, and
showed that the dark energy itself never collapses to form black holes,  
but when both the dark energy and the dust are present, black holes 
can be formed, due to the condensation of the dust. %The results do not
%support the speculations that black holes cannot exist at all.
%%%%%%%%%%%%%%%%%%%%%%%%%%%%%%%%%%%%%%%%%%%%%%%%%%%%%%%%%%%%%%%%%%%%%%%
%But \cite{5} do not consider the junction of the star to the spacetime
%outside of it.
%%%%%%%%%%%%%%%%%%%%%%%%%%%%%%%%%%%%%%%%%%%%%%%%%%%%%%%%%%%%%%%%%%%%%%%
Similar results were obtained in \cite{6}. Of course, cases of the
most interesting are black holes in the background of our real
universe. In 1933, McVittie \cite{7} found the exact solution of
Einstein's equations for a perfect fluid,
which in general describes   a Schwarzschild black hole being
embedded in a Friedmann-Robertson-Walker (FRW) universe. The
solution was extended to the cases of charged black holes \cite{8}
and arbitrary dimensions \cite{9}, and the global structure of the
solution was also investigated in some detail \cite{10,11}. Recall
that dark energy usually violates the strong energy condition
\cite{12}.

In this paper, we are going to re-examine McVittie's solutions and
show that black holes can exist even in the background of a dark energy dominated
expanding universe. Specifically, in the next section, we review McVittie's
solutions and consider  three special cases. 
In Sec. III, we give a proof for
the existence of black holes in such backgrounds, while in Sec. IV
we study the evolution of the apparent horizons of the asymptotic
Schwarzschild, singular, and Schwarzschild-de Sitter models,
respectively. We conclude the paper with some comments in Sec. V.

\section{ McVittie's Solutions with Dark Energy}

\renewcommand{\theequation}{2.\arabic{equation}}
\setcounter{equation}{0}

\subsection{McVittie's Solutions}

McVittie's solutions can be written in the form \cite{7,11}
\bq
\lb{2.1}
ds^{2} = -\left(\frac{1-\frac{M}{2N}}{1+\frac{M}{2N}}\right)^{2}dt^{2}
+ e^{\beta}\left(1+\frac{M}{2N}\right)^{4}(dr^{2}+ h^{2}d\Omega^{2}),
\eq
where $d\Omega^{2} \equiv d\theta^{2} + \sin^{2}\theta d\phi^{2}$, and
\bq
\lb{2.2}
\beta=\beta(t), \;\;\; M=m e^{-\frac{\beta}{2}},
\eq
with $m$ being the mass parameter. The functions $h(r)$ and $N(r)$ depend on 
the choice of the constant $k$, and are given, respectively, by
\bqn
\lb{2.3}
h(r) & =&
\cases{\sinh r, & $k = -1$,\cr r, & $k = 0$,\cr
\sin r, & $k = +1$,\cr}\nb\\
N(r) & =& \cases{2 \sinh \frac{r}{2}, & $k = -1$,\cr
r, & $k = 0$,\cr
2\sin\frac{r}{2}, & $k = +1$.\cr}
\eqn

In 1994, Nolan \cite{10,11}  studied the problem in some details, and
 gave the criteria for a point mass
embedded in an open FRW universe. He showed that McVittie's solution in
the case $k=0$ satisfies the criteria, but does not in the case
$k=1$. Therefore, we consider only the case $k=0$ in present paper.
In this case, the metric may be rewritten as
\bq
\lb{2.4}
ds^{2} =
-\left(\frac{1-\frac{m}{2u}}{1+\frac{m}{2u}}\right)^{2}dt^{2} +
e^{\beta}\left(1+\frac{m}{2u}\right)^{4}(dr^{2}+ r^{2}d\Omega^{2}),
\eq
where
\bq
\lb{2.5}
u \equiv r e^{\frac{\beta}{2}}.
\eq
Clearly, when $m = 0$ the McVittie solutions reduce to the FRW one
with its expansion factor  given by
\bq
\lb{2.4a}
a(t) = e^{\beta(t)/2}.
\eq

Making the transformation \cite{10} 
\bq 
\lb{2.6}
R=u(1+\frac{m}{2u})^{2}, 
\eq 
or inversely, 
\bq 
\lb{2.6a} 
u = \frac{1}{2}\left[(R-m) - \epsilon\sqrt{R(R-2m)}\right], 
\eq 
we find that 
\bqn 
\lb{2.7} 
  e^{\beta}\left(1+\frac{m}{2u}\right)^{4}dr^{2} &=& \frac{1}{4}
\dot{\beta}^{2}R^{2}dt^{2} \nb\\
& & + \epsilon\dot{\beta}R
    \left(1-\frac{2m}{R}\right)^{-\frac{1}{2}}dtdR\nb\\
& & + \left(1-\frac{2m}{R}\right)^{-1}dR^{2},\\
\lb{2.8}
\frac{m - 2u}{m + 2u} &=&    \epsilon \sqrt{1-\frac{2m}{R}},
\eqn
where   $\epsilon = \pm 1$. From the above expressions, we can see that
$\epsilon$ must be chosen such that,
\bq
\lb{2.10}
 \;\;\;
\epsilon = \cases{+ 1, & $u\in(0,\frac{m}{2})$, \cr
 - 1, & $ u\in(\frac{m}{2},\infty)$,\cr}
\eq
for   $m \neq 0$; and
\bq
\lb{2.11}
\epsilon=-1,
\eq
for $m = 0$. Note that the radius $R$ varies 
over the interval $ R \in (2m, \infty)$.
As shown below, the spacetime usually is singular at $R = 2m$.

Substituting Eqs.(\ref{2.7}) and (\ref{2.8}) 
into Eq. (\ref{2.4}), we obtain
\bqn
\lb{2.9} ds^{2} &=&
-\left(1-\frac{2m}{R}-\frac{1}{4}\dot{\beta}^{2}R^{2}\right)dt^{2}\nb\\
& &
+   \epsilon\dot{\beta}R\left(1-\frac{2m}{R}\right)^{-\frac{1}{2}}dR
dt \nb\\
& & + \left(1-\frac{2m}{R}\right)^{-1}dR^{2}+ R^{2}d\Omega^{2},
\eqn
where the dot denotes the derivative with respect to $t$.

From Einstein's field equations one can obtain the corresponding
energy density and isotropic pressure,
\bq
\lb{2.12}
8\pi\rho=\frac{3}{4}\dot{\beta}^{2}, \;\;\; 8\pi
p=-\frac{3}{4}\dot{\beta}^{2}
+   \epsilon\ddot{\beta}\left(1-\frac{2m}{R}\right)^{-\frac{1}{2}}.
\eq
 Obviously, $R=2m$ is an intrinsic spacetime singularity for
$\ddot{\beta} \not= 0$. 
%It is not possible to extend the spacetime
%across this singularity in a continuous manner.
%%%%%%%%%%%%%%%%%%%%%%%%
%Therefore, our
%spacetime is defined within $2m < R < \infty.$

From Eq.(\ref{2.12}) we can see that in the present case  the energy density is
always non-negative. Then, the weak energy condition reduces to
\bq
\lb{2.13}
\rho + p\geq0.
\eq
Equation (\ref{2.12}) shows that the
condition is satisfied when $\epsilon\ddot{\beta}\geq 0$.

\subsection{Special Solutions of $m=0$}

The energy density and pressure of the FRW background are given by
setting $m=0$ in Eq.(\ref{2.12}), which read
\bq
\lb{2.14}
8\pi\rho_{0}=\frac{3}{4}\dot{\beta}^{2}, \;\;\; 
8\pi p_{0}=-\frac{3}{4}\dot{\beta}^{2}-\ddot{\beta}.
\eq
In writing the above expressions,  following Eq.(\ref{2.11}) we had 
set $\epsilon = -1$. As already mentioned in the introduction, 
the dark energy satisfies the condition
$\rho_{0}+3p_{0}<0$. Setting
\bq
\lb{2.15}
\frac{1}{2}\dot{\beta}^{2}+\ddot{\beta} \equiv F^{2}(t),
\eq
one can show that the background is always filled with dark energy,
 where $ F(t)$ is a real and otherwise arbitrary function. For
simplicity, we choose F(t) such that
\bq
\lb{2.16}
F^{2}(t)=c \dot{\beta}^{2}+\delta,
\eq
where $c$ and $\delta$ are positive constants. Then, Eq.(\ref{2.15})
has the first integral,
\bq
\lb{2.17}
\dot{\beta}^{2}=\alpha+\gamma e^{-2\frac{\beta}{b}},
\eq
where
\bq
\lb{2.18}
\alpha=b \delta, \;\;\; b=\frac{2}{1-2c},
\eq
and $\gamma$ is a real constant. To solve Eq.(\ref{2.17}),
 it is found convenient to
distinguish the three cases: $\alpha=0$, $\alpha<0$ and $\alpha>0$.

\subsubsection{ $\alpha=0$}

 In this case, we have $\delta=0$, and  $b$ can be
written as
\bq
\lb{2.19}
b=\frac{4}{3(1+w)}.
\eq
Then, the equation of state reads
\bq
\lb{2.20}
p_{0}=w\rho_{0},
\eq
which tells us that the FRW background is linear. It can easily be shown
that the solution is given by
\bq
\lb{2.21}
\dot{\beta}=\frac{b}{t}, \;\;\;\ddot{\beta}=-\frac{b}{t^{2}}.
\eq
Although Eq.(\ref{2.21}) has the
same forms as that of \cite{10}, the value of $b$ is different. For
dark energy, the region $w\in(-\frac{1}{3}, -1)$ corresponds to the
region $b\in(2, \infty)$, while the region $w<-1$ corresponds to
$b<0$. Obviously, for $\epsilon=-1$ the weak energy condition is
satisfied when  $w\in(-\frac{1}{3}, -1)$; and for $\epsilon=1$ it
is satisfied when  $w<-1$.

\subsubsection{ $\alpha < 0$}

In this case, we find that $b<0$ and $c>\frac{1}{2}$. 
The solution is given by
\bq
\lb{2.22}
\dot{\beta}=\pm\sqrt{-\alpha}\cot\left(\frac{\sqrt{-\alpha}}{ b}t\right),
\;\;\;
\ddot{\beta}=\pm\frac{\alpha}{b}\csc^{2}\left(\frac{\sqrt{-\alpha}}{
b}t\right).
\eq
From Eqs.(\ref{2.22}), it is clear that, when $\epsilon=1$, the
weak energy condition is satisfied for
$\dot{\beta}=\sqrt{-\alpha}\cot(\frac{\sqrt{-\alpha}}{ b}t)$; and when
$\epsilon=-1$, the weak energy condition is satisfied for
$\dot{\beta}=-\sqrt{-\alpha}\cot(\frac{\sqrt{-\alpha}}{ b}t)$.

\subsubsection{ $\alpha > 0$}

When $\alpha>0$, we have $b\in(2, \infty)$ and
$c\in( 0, \frac{1}{2} )$. The solution then is given by
\bq
\lb{2.23}
\dot{\beta}=\pm\sqrt{\alpha}\coth\left(\frac{\sqrt{\alpha}}{
b}t\right), \;\;\;
\ddot{\beta}=\mp\frac{\alpha}{b}\left[\sinh\left(\frac{\sqrt{\alpha}}{
b}t\right)\right]^{-2}.
\eq
Eq.(\ref{2.23}) shows that, when $\epsilon=1$ the weak
energy condition holds for
$\dot{\beta}=-\sqrt{\alpha}\coth(\frac{\sqrt{\alpha}}{ b}t)$; and when
$\epsilon=-1$ the weak energy condition holds for
$\dot{\beta}=\sqrt{\alpha}\coth(\frac{\sqrt{\alpha}}{ b}t)$.

Note that  all previous discussions of the weak energy condition were
based on the cases where the interaction of matter and dark energy
exist, that is $m \not= 0$. From Eq.(\ref{2.14}) we can see that,
for a dark energy dominated background, the
weak energy condition is satisfied if and only if $\ddot{\beta}<0$.  

\section{Existence of Black Holes}

\renewcommand{\theequation}{3.\arabic{equation}}
\setcounter{equation}{0}

In the previous section, we have derived three special solutions in
a homogeneous and isotropic FRW universe with dark energy. The corresponding
function $\beta(t)$ is given, respectively, by Eqs.(\ref{2.21}), (\ref{2.22}) 
and (\ref{2.23}). let us now show that, when $m \not= 0$, these solutions 
represent black holes in the background of a FRW universe  filled with
dark energy.

According to \cite{13,14,15}, black holes are defined by the
existence of {\em future outer apparent horizons}. To study apparent
horizons, we introduce two null coordinates $\xi^{+}$ and $\xi^{-}$
via the relations
\bqn
\lb{3.1}
  d\xi^{+} &=& f\left\{\left(1-\frac{2m}{R}\right)^{\frac{1}{2}}dt +
\frac{1}{2}\epsilon\dot{\beta}R dt \right. \nb\\
& & \;\;\;\;\;\; \left. +
\left(1-\frac{2m}{R}\right)^{-\frac{1}{2}}dR\right\},\nb\\
% \lb{3.1}
  d\xi^{-} &=& g\left\{\left(1-\frac{2m}{R}\right)^{\frac{1}{2}}dt -
\frac{1}{2}\epsilon\dot{\beta}R dt \right.\nb\\
& & \;\;\;\;\;\; \left. -
\left(1-\frac{2m}{R}\right)^{-\frac{1}{2}}dR\right\}, 
\eqn
where $f$ and $g$ are functions of $t$ and $R$ only, 
and satisfy the  integrability conditions,
\bq
\lb{3.1a}
\frac{\partial^{2}\xi^{\pm}}{\partial t \partial R} =
\frac{\partial^{2}\xi^{\pm}}{\partial R \partial t}.
\eq
Without loss of generality, we assume  that they are positive,
\bq
\lb{3.1b}
f > 0, \;\;\; g > 0.
\eq
Then, it can be shown that both $\xi^{+}$ and $\xi^{-}$ are future-pointing, and
along the lines of constant $\xi^{-}$ the radial coordinate $R$ increase towards the
future, while along the lines of constant $\xi^{+}$ the coordinate $R$ decreases
towards the future. In terms of $\xi^{\pm}$,  the metric (\ref{2.9}) can be written as
\bq
\lb{3.2}
ds^{2}=-2 e^{2\sigma(\xi^{+}, \xi^{-})}d\xi^{+}d\xi^{-}
+ R^{2}(\xi^{+}, \xi^{-}) d\Omega^{2},
\eq
where
\bq
\lb{3.3}
 \sigma(\xi^{+}, \xi^{-}) \equiv -\frac{1}{2}\ln(2fg).
\eq

Introducing the two  null  vectors
$\xi_{(\pm)} \equiv \frac{\partial}{\partial\xi^{\pm}}$, which are
future-pointing, we find that $\xi_{(\pm)}$ define two null geodesic congruences,
\bq
\lb{3.4a}
\xi_{(\pm);\nu}^{\mu} \xi_{(\pm)}^{\nu} = 0,
\eq
and the expansions of these   congruences are given by
\bqn
\lb{3.4}
\theta_{+} &\equiv& \xi_{(+)\mu ;\nu} g^{\mu\nu}\nb\\
&=& \frac{1}{Rf}
\left[\left(1-\frac{2m}{R}\right)^{\frac{1}{2}}-
\frac{1}{2}\epsilon\dot{\beta}R\right],\\
\lb{3.5}
\theta_{-}&\equiv & \xi_{(-)\mu ;\nu} g^{\mu\nu}\nb\\
&=&
-\frac{1}{Rg} \left[\left(1-\frac{2m}{R}\right)^{\frac{1}{2}}+\frac{1}{2}
\epsilon\dot{\beta}R\right].
\eqn

Following  \cite{13,14,15}, we define that {\em  a
two sphere, $\cal{S} $, of constant $t$ and $R$, is said to be trapped
if $\theta_{+}\theta_{-}>0$, untrapped if
$\theta_{+}\theta_{-}<0$, and marginal if
$\theta_{+}\theta_{-}=0$}.

Assuming that on the marginally trapped
surfaces $\cal{S}$ we have $\left.\theta_{+}\right|_{\cal{S} } = 0$, then an
{\em apparent horizon} is
the closure $\tilde{\Sigma}$ of a three-surface $\Sigma$ foliated by the trapped
surfaces $\cal{S} $ on which $\left.\theta_{-}\right|_{\Sigma} \not= 0$.
It is said {\em outer, degenerate, or inner}, according to
whether $\left.{\cal{L}}_{-}\theta_{+}\right|_{\Sigma} < 0$,
$\left.{\cal{L}}_{-}\theta_{+}\right|_{\Sigma} = 0$, or
$\left.{\cal{L}}_{-}\theta_{+}\right|_{\Sigma} > 0$, , where ${\cal{L}}_{-} \equiv
{\cal{L}}_{\xi_{(-)}}$ denotes the Lie derivative along
$\xi_{(-)}$. In addition, if $\left. \theta_{-}\right|_{\Sigma} < 0$
then the apparent horizon is said {\em future}, and if
$\left. \theta_{-}\right|_{\Sigma} > 0$ it is said {\em past}.

{\em Black holes} are usually defined by the existence of {\em future outer
apparent horizons} \cite{13,14,15}. However, in a definition given by
Tipler \cite{Tip77} the degenerate case was also included \cite{13}.

From Eqs.(\ref{3.4}) and (\ref{3.5}), we find that on the   apparent
horizons we have, 
\bq 
\lb{3.6}
\left. \left(1-\frac{2m}{R}-\frac{1}{4}\dot{\beta}^{2}R^{2}\right)\right|_{AH}=0. 
\eq 
Thus, so long as 
\bq 
\lb{3.7} 
\dot{\beta}^{2}<\frac{4}{27 m^{2}}, 
\eq
Eq.(\ref{3.6}) has two solutions, $R = R_{H}$ and $R = R_{C}$, where
\bq
\lb{3.7aa}
R_{H}< 3m <R_{C}. 
\eq
It can also be  shown that 
\bqn 
\lb{3.8}
{\cal{L}}_{-} \theta_{+}|_{AH} &=&
\frac{1}{4fg}\left\{\frac{4}{R^{2}}\left(1-\frac{3m}{R}\right)\right.\nb\\
& & \left. -
\epsilon\ddot{\beta}\left(1-\frac{2m}{R}\right)^{-\frac{1}{2}}\right\}. 
\eqn

Since black holes are defined by  the existence of future outer
apparent horizons, we must have $\theta_{+}|_{AH}=0$,
$\theta_{-}|_{AH}<0$, and ${\cal{L}}_{-} \theta_{+}|_{AH}<0$. From
Eqs. (\ref{3.4}), (\ref{3.5}) and (\ref{3.8}), 
it is clear that, if and only if
\bq
\lb{3.9}
\epsilon\dot{\beta}>0,
\eq
and
\bq
\lb{3.10}
\epsilon\ddot{\beta}>\frac{4}{R^{2}}\left(1-\frac{3m}{R}\right)
\left(1-\frac{2m}{R}\right)^{\frac{1}{2}},
\eq
black holes exist in the FRW universe. Note that these conditions not
only apply to the dark energy dominated universe, but also to the
matter-dominated universe.  For $\epsilon=-1$, we find
that,  even there exist  apparent horizons, they
cannot be  future outer apparent horizons in an expanding universe.
This is consistent with the result obtained in  \cite{10}.

Now let us show that, for $\epsilon=1$, solutions satisfying these
conditions indeed exist in an expanding universe $\dot{\beta} > 0$.
From Eq.(\ref{3.6}), on the hypersurface $R = R_{H}$ we have 
\bq 
\lb{3.11}
\dot{\beta} =\frac{2}{R_{H}}\left(1-\frac{2m}{R_{H}}\right)^{\frac{1}{2}}. 
\eq
Then, we find 
\bq 
\lb{3.12}
\ddot{\beta}=-\frac{\dot{R}_{H}}{2\left(1-\frac{2m}{R_{H}}\right)}
\left[\frac{4}{R^{2}_{H}}\left(1-\frac{3m}{R_{H}}\right)
\left(1-\frac{2m}{R_{H}}\right)^{\frac{1}{2}}\right].
\eq 
In all the three cases, $\alpha = 0, \; \alpha < 0$, and $\alpha
> 0$, one can show that $\dot{R}_{H}<0$. Then, 
considering Eq.(\ref{3.12}) and the fact that $1-\frac{3m}{R_{H}}<0$,
we find that the condition (\ref{3.10}) requires 
\bq 
\lb{3.13} I_{H}
\equiv 2\left(1-\frac{2m}{R_{H}}\right)+\dot{R}_{H}>0.
\eq 
 
%%%%%%%%%%%%%%%%%%%%%%%%%%%%%%%%%%%%%%%%%%%%%%%%%%%%%%%%%%%%%%%%%%%%%%%%%%%%%%
\begin{figure}
\includegraphics{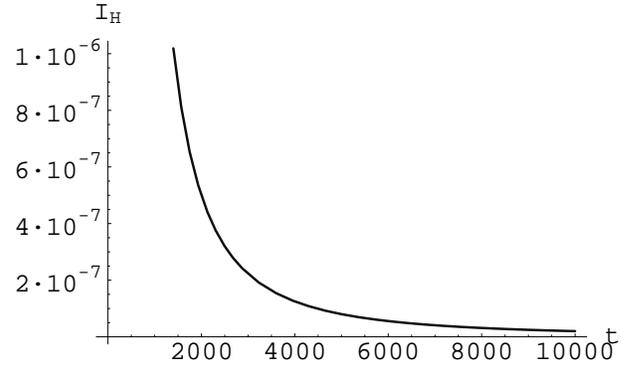}
\caption{The function $I_{H}$ for the case $\alpha = 0$ with $m=1$ and
$b=1$.}\label{fig1}
\end{figure}
%%%%%%%%%%%%%%%%%%%%%%%%%%%%%%%%%%%%%%%%%%%%%%%%%%%%%%%%%%%%%%%%%%%%%%%%%%%%%%%

As shown in Fig. 1 and Fig. 2,  solutions satisfying Eq.(\ref{3.13}) indeed
exist in the   cases, $\alpha = 0 $ and $ \alpha < 0$. For the case $ \alpha > 0$,
Eq.(\ref{3.13})  is obviously satisfied, since when $t\rightarrow\infty$, one has
$R_{H}>2m$ and $\dot{R}_{H}\rightarrow 0$. Therefore, we conclude that
black holes exist even in dark energy dominated universe. It should
be noted that $R_{C}$ may be interpreted as the location of the
cosmological apparent horizon.

%%%%%%%%%%%%%%%%%%%%%%%%%%%%%%%%%%%%%%%%%%%%%%%%%%%%%%%%%%%%%%%%%%%%%%%%%%%%%%
\begin{figure}
\includegraphics{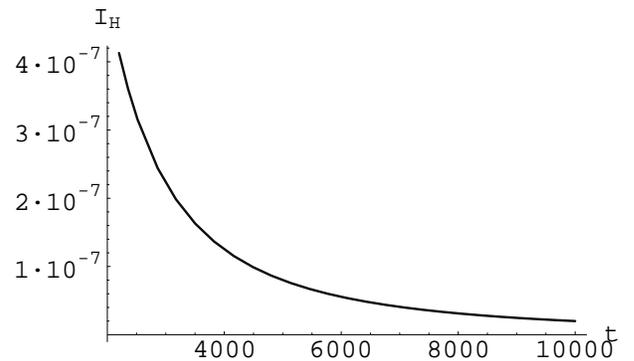}
\caption{The function $I_{H}$ for the case $\alpha < 0$ with $m=1$, $b=-1$
and $\sqrt{-\alpha}=10^{-10}$.}\label{fig2}
\end{figure}
%%%%%%%%%%%%%%%%%%%%%%%%%%%%%%%%%%%%%%%%%%%%%%%%%%%%%%%%%%%%%%%%%%%%%%%%%%%%%%%

\section{Evolution of apparent Horizons}

\renewcommand{\theequation}{4.\arabic{equation}}
\setcounter{equation}{0}

In this section, we consider the evolution of the apparent horizons
for $\alpha=0$, $\alpha<0$ and $\alpha>0$, respectively. All of
these models have a big bang singularity. From Eq.(\ref{3.6}), we find
that the apparent
horizons $R_{H}$ and $R_{C}$ are given by
\bqn
\lb{4.1}
R_{H}&=&\frac{4}{\sqrt{3}\mid
\dot{\beta}\mid}\cos\left(\frac{\Psi}{3}+\frac{\pi}{3}\right),\\
\lb{4.2} R_{C} &=& \frac{4}{\sqrt{3}\mid
\dot{\beta}\mid}\cos\left(\frac{\Psi}{3}-\frac{\pi}{3}\right), 
\eqn 
where 
\bq
\lb{4.3} 
\Psi=\arccos\left(\frac{3\sqrt{3}}{2}m\mid\dot{\beta}\mid\right). 
\eq
Since we are dealing with  black holes   in an expanding FRW universe filled
with dark energy, in the following discussions we consider only the case  
$\dot{\beta}> 0$ and $\epsilon = 1$.

\subsection{$\alpha=0$}

In this case, the function $\beta(t)$ is given by Eq.(\ref{2.21}), 
from which we
see that the spacetime is singular at $t=0$, which may be
interpreted as the big bang singularity \cite{10}.

As shown in Fig. 3, the two apparent horizons appear at the same
moment $t= t_{0} \equiv \frac{3\sqrt{3}}{2}mb$. As time increases,
$R_{H}$ becomes smaller and smaller, while $R_{c}$ becomes larger
and larger. When $t\rightarrow\infty$, we have $R_{H}\rightarrow 2m$
and $R_{C}\rightarrow\infty$. Therefore, this is the asymptotic
Schwarzschild  model.
%Note that, for this case, $R=2m$ is not a
%singularity of infinite pressure at $t\rightarrow \infty$.

%%%%%%%%%%%%%%%%%%%%%%%%%%%%%%%%%%%%%%%%%%%%%%%%%%%%%%%%%%%%%%%%%%%%%%%%%%%%%%
\begin{figure}
\includegraphics{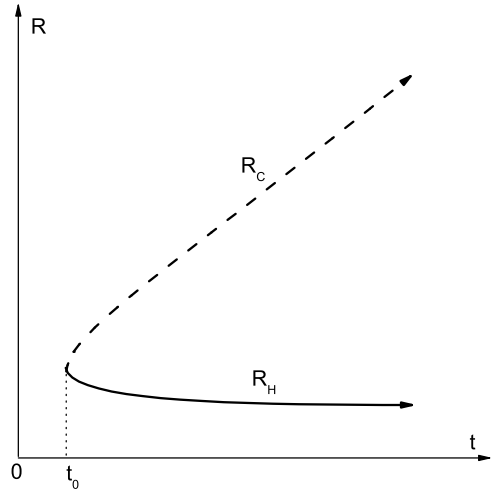}
\caption{The evolution of the apparent horizons for the asymptotic
Schwarzschild model ($\alpha=0$). The apparent horizons first
develop at the moment $t=t_{0}
\equiv\frac{3\sqrt{3}}{2}mb$.}\label{fig3}
\end{figure}
%%%%%%%%%%%%%%%%%%%%%%%%%%%%%%%%%%%%%%%%%%%%%%%%%%%%%%%%%%%%%%%%%%%%%%%%%%%%%%%

\subsection{$\alpha<0$}

In this model, our spacetime can be defined within $0\leq
t\leq-\frac{b \pi}{2 \sqrt{-\alpha}}$. The evolution of the apparent
horizons with time are shown in Fig. 4, where $t=0$ is a big bang
singularity. The apparent horizons first develop at
$t=t_{0}=-\frac{b}{\sqrt{-\alpha}}\cot^{-1}(\frac{2}{3m\sqrt{-3\alpha}})$
and immediately bifurcate. As time increases, $R_{H}$ decreases
monotonically, while $R_{c}$ increases monotonically. When
$t=t_{s}=-\frac{b\pi}{2\sqrt{-\alpha}}$, the apparent horizon
$R_{H}$ coincides with the singularity at $R=2m$, and $R_{C}$
expands to infinity. This is the asymptotically singular model. Note
that, if our spacetime is defined within $-\frac{b \pi}{2
\sqrt{-\alpha}}\leq t\leq-\frac{b \pi}{\sqrt{-\alpha}}$, the
evolution of the apparent horizons is similar to the time reversal
of the previous case.

%%%%%%%%%%%%%%%%%%%%%%%%%%%%%%%%%%%%%%%%%%%%%%%%%%%%%%%%%%%%%%%%%%%%%%%%%%%%%%
\begin{figure}
\includegraphics{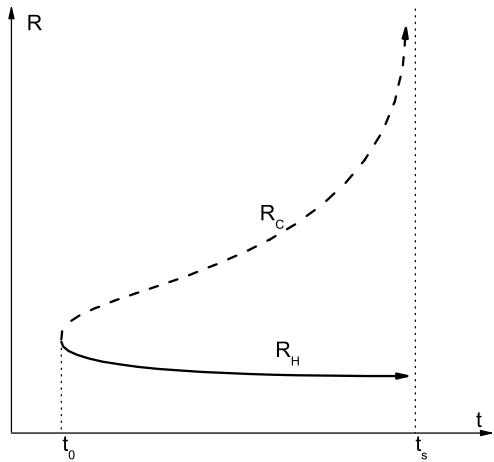}
\caption{The evolution of the apparent horizons for the
asymptotically singular  model ($\alpha<0$). The apparent horizons
develop at the moment $t=t_{0} \equiv
-\frac{b}{\sqrt{-\alpha}}\cot^{-1} (\frac{2}{3m\sqrt{-3\alpha}})$.}
\label{fig4}
\end{figure}
%%%%%%%%%%%%%%%%%%%%%%%%%%%%%%%%%%%%%%%%%%%%%%%%%%%%%%%%%%%%%%%%%%%%%%%%%%%%%%%

\subsection{$\alpha>0$}

In this case, the functin $\beta(t)$ is that of Eq.(\ref{2.23}). As shown in Fig. 5,
the apparent horizons first develop at
$t=\frac{b}{\sqrt{\alpha}}\coth^{-1}(\frac{2}{3m\sqrt{3\alpha}})$
and immediately bifurcate. As time increases, $R_{H}$ decreases, and
 $R_{C}$ increases. When $t\rightarrow\infty$,
they tend asymptotically to two different constants.
Consequently, the model is  asymptotically Schwarzschild-de Sitter.

%%%%%%%%%%%%%%%%%%%%%%%%%%%%%%%%%%%%%%%%%%%%%%%%%%%%%%%%%%%%%%%%%%%%%%%%%%%%%%
\begin{figure}
\includegraphics{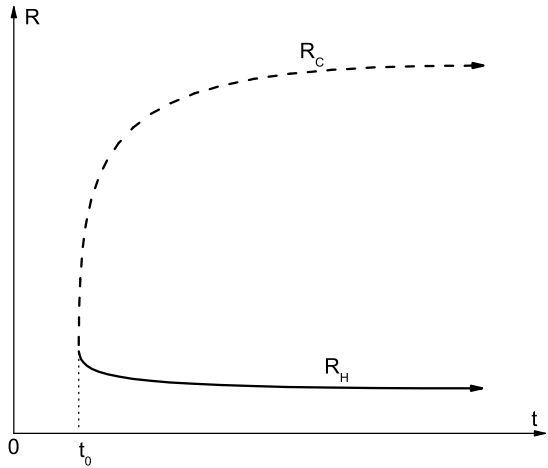}
\caption{The evolution of the apparent horizons for the asymptotic
Schwarzschild-de Sitter model ($\alpha>0$). The apparent horizons
develop at the moment
$t=t_{0}\equiv\frac{b}{\sqrt{\alpha}}\coth^{-1}(\frac{2}{3m\sqrt{3\alpha}}).$
}\label{fig5I}
\end{figure}
%%%%%%%%%%%%%%%%%%%%%%%%%%%%%%%%%%%%%%%%%%%%%%%%%%%%%%%%%%%%%%%%%%%%%%%%%%%%%%%

\section{Conclusions}

\renewcommand{\theequation}{5.\arabic{equation}}
\setcounter{equation}{0}

We have investigated the issue of the existence of black hole in
the background of a homogeneous and isotropic expanding FRW universe filled 
with dark energy. We have shown that, when $\epsilon = +1$, and the conditions
\bqn
\lb{5.1}
& & \dot{\beta}>0,\\
\lb{5.2}
& & \ddot{\beta}>\frac{4}{R^{2}}\left(1-\frac{3m}{R}\right)
\left(1-\frac{2m}{R}\right)^{\frac{1}{2}},
\eqn
are satisfied,  black holes  in a dark energy
dominated expanding FRW background exist.  
It is interesting to note that the background of these solutions
satisfies the weak energy condition.  Therefore, the speculations 
that black holes do not exist in our universe due 
to the presence of dark energy is groundless.

To show the above claim explicitly,   we have considered  three
special McWittie solutions, which are  the asymptotically
Schwarzschild, asymptotically singular,  and asymptotically
Schwarzschild-de Sitter solutions, given, respectively, by  
Eqs.(\ref{2.21})-(\ref{2.23}) with  $\dot{\beta}> 0$. 
For the asymptotic Schwarzschild solution, the FRW background is linear. 
For the asymptotically singular solution, the time of evolution of
universe is finite. For the asymptotic Schwarzschild-de Sitter
solution, the final form of the equation of state of the FRW
background is $w = -1$.

\section*{ACKNOWLEDGMENTS}

One of the authors (ZHL) would like to thank Physics Department, Baylor
University  for hospitality. AW would like to thank the Center of Astrophysics,
Zhejiang University of Technology for hospitality, and Baylor University
for Summer 2006 sabbatical leave.  This work was supported partly by the National
Natural Science Foundation of China under Grant No. 10375051 (ZHL).

%%%%%%%%%%%%%%%%%%%%%%%%%%%%%%%%%%%%%%%%%%%%%%%%%%%%%%%%%%%%%%%%%%%%%%%%%%%5

\end{document}